\documentclass[11pt]{article}
\usepackage{graphics,amssymb}
\usepackage{epsfig}
\begin{document}

{
\setlength{\textwidth}{16.5cm}
\setlength{\textheight}{22.2cm}
\setlength{\hoffset}{-1.43cm}
\setlength{\voffset}{-.9in}

\thispagestyle{empty}
\renewcommand{\thefootnote}{\fnsymbol{footnote}}

\begin{flushright}
{\normalsize
SLAC-AP-128\\
July 2000}
\end{flushright}

\vspace{.8cm}

\begin{center}
{\bf\Large Obtaining the
 Wakefield Due to Cell-to-Cell Misalignments in a Linear Accelerator
Structure
\footnote{\small Work supported by
Department of Energy contract  DE--AC03--76SF00515.}}

\vspace{1cm}

{\large
Karl L.F. Bane and Zenghai Li\\
Stanford Linear Accelerator Center, Stanford University,
Stanford, CA  94309}

\end{center}
}
\vfill

\def\la{\langle} 
\def\ra{\rangle} 
\def\lm{\lambda}

\title{Obtaining the
 Wakefield Due to Cell-to-Cell Misalignments in a Linear Accelerator
Structure}
\author{Karl L.F. Bane and Zenghai Li}
\date{}
\maketitle

We are interested in obtaining the long-range, dipole wakefield
of a linac structure with internal misalignments. 
The NLC linac structure is composed of a collection of cups
that are brazed together, and such a calculation, for example,
is important in setting the straightness 
tolerance for the composite structure. 
Our derivation, presented here, is technically applicable only to structures
for which all modes are trapped. The modes will be trapped at least at the
ends of the structure, if the connecting beam tubes have sufficiently
small radii and the dipole modes do not couple to the fundamental
mode couplers in the end cells.
 For detuned structures (DS), like those in the injector
linacs of the JLC/NLC\cite{BL}, most modes are trapped internally within
a structure, and those that do extend to the ends couple only weakly
to the beam; for such structures the results here can also be applied,
even if the conditions on the beam tube radii and the fundamental mode
coupler do not hold.
We believe that even for the damped, detuned structures (DDS) of the
main linac of the JLC/NLC\cite{Jones}, which are similar, though they have
manifolds to add weak damping to the wakefield, a result
very similar to that presented here applies.

\begin{figure}[htb]
\centering
\epsfig{file=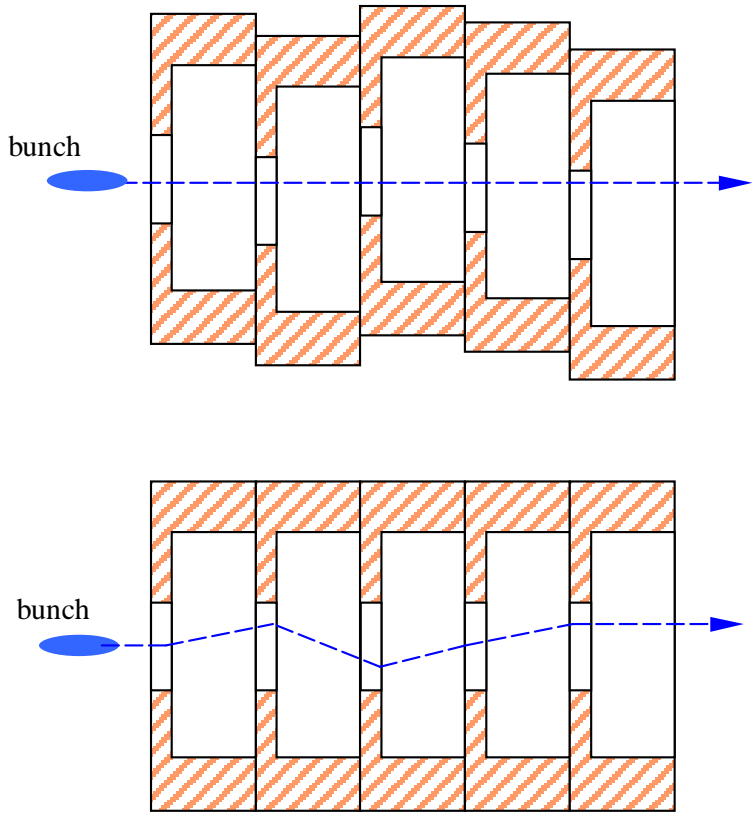, width=10cm}
\caption{
Sketches of part of a misaligned structure (top) and the model used
for the kick factor calculation (bottom). 
Note that the relative size of the misalignments here is much exaggerated.
}
\label{fistruct}
\end{figure}

We assume a structure is composed of many cups that are misaligned 
transversely by
amounts that are very small compared to the cell dimensions.
For such a case we assume that the mode frequencies are the same as in
the ideal structure,
and only the mode kick factors are affected. 
To first order we assume that for each mode, the kick factor for the beam
on-axis in the imperfect structure is the same as for the case with the beam
following the negative of the misalignment path in the error-free structure.
In Fig.~\ref{fistruct} we sketch a portion of such a misaligned structure (top)
and the model used for the kick factor calculation (bottom).
Note that the relative size of the 
misalignments is exaggerated from what is expected, in order
to more clearly show the principle.
Given this model,
the method of calculation of the kick factors 
can be derived using the so-called 
``Condon Method''\cite{condon},\cite{morton} (see also \cite{bane}). 
Note that this 
application to cell-to-cell misalignments in an accelerator structure
is presented in Ref.~\cite{perturb}. 
The results of this perturbation method have been shown to be consistent
with those using a 3-dimensional scattering matrix analysis\cite{Valery}.
We will only sketch the derivation below.

Consider a closed cavity with perfectly conducting walls.
For such a cavity the Condon method 
expands the vector and scalar potentials,
in the Coulomb gauge, as a sum over the empty cavity modes.
As function of position $\bf x$ $(x,y,z)$ and time $t$ 
the vector potential in the cavity is given as
\begin{equation}
{\bf A}({\bf x},t)=\sum_\lm q_\lm(t){\bf a}_\lm({\bf x})\quad,
\end{equation}
where
\begin{equation}
\nabla^2{\bf a}_\lm + {\omega_\lm^2\over c^2}{\bf a}_\lm=0\quad,
\end{equation}
with $\omega_\lm$ the frequency of mode $\lm$, and
${\bf a}_\lm\times{\bf{\hat n}}=0$
on the metallic surface (${\bf{\hat n}}$ is a unit vector normal
to the surface).
Using the Coulomb gauge implies that $\nabla\cdot{\bf a}_\lm=0$.
The $q_\lm$ are given by
\begin{equation}
\ddot{q}_\lm+\omega_\lm^2 q_\lm={1\over 2U_\lm}\int_{\cal{V}}d\cal{V}\,
  {\bf j}\cdot {\bf a}_\lm\quad,
\end{equation}
with the normalization
\begin{equation}
{\epsilon_0\over2}\int_{\cal{V}}d\cal{V}\,
 {\bf a}_{\lm^\prime}\cdot{\bf a}_\lm= U_\lm\delta_{\lm\lm^\prime}\quad,
\end{equation}
with ${\bf j}$ the current density. Note that
the integrations are performed over the volume of the cavity $\cal{V}$.

The scalar potential is given as
\begin{equation}
\Phi({\bf x},t)=\sum_\lm r_\lm(t)\phi_\lm({\bf x})\quad,
\end{equation}
where
\begin{equation}
\nabla^2\phi_\lm + {\Omega_\lm^2\over c^2}\phi_\lm=0\quad,
\end{equation}
with $\Omega_\lm$ the frequencies associated
with $\phi_\lm$, and with $\phi_\lm=0$ on the metallic surface.
The $r_\lm$ are given by
\begin{equation}
r_\lm={1\over 2T_\lm}\int_{\cal{V}}d\cal{V}\,\rho\phi_\lm\quad,
\end{equation}
with $\rho$ the charge distribution in the cavity.
Note that one fundamental difference between the behavior
of $A({\bf x},t)$ and $\Phi({\bf x},t)$ is that
when there are no charges in the cavity the vector potential
can still 
oscillate whereas the scalar potential must be identically equal to 0.

Let us consider an ultra-relativistic driving charge $Q$ that passes through
the cavity parallel to the $z$ axis, and (for simplicity) a test charge
following at a distance $s$ behind on the same path.
 Both enter the cavity at position $z=0$
and leave at position $z=L$.
The transverse wakefield at the test charge is then
\begin{eqnarray}
{\bf W}(s)&=&{1\over QLx_0}\int_0^L dz\,\left[c\nabla_\bot A_z-\nabla_\bot\Phi\right]_{t=(z+s)/c}
            \nonumber\\
          &=&{1\over QLx_0}\sum_\lm\int_0^L dz\,\left[cq_\lm\left({z+s\over c}\right)\nabla_\bot
          a_{\lm z}(z)\right.\nonumber\\
          & &\quad\quad\quad\quad\quad\quad\quad\quad\quad
          -r_\lm\left({z+s\over c}\right)\nabla_\bot\phi_\lm(z)\bigg]\ ,
\label{eqconda}
\end{eqnarray}
where the integrals are along the path of the particle trajectory.
The parameter $x_0$ is a parameter for transverse offset
(the transverse wake is usually given in units of V/C per longitudinal
meter per transverse meter);
for a cylindrically-symmetric structure it is usually taken to be the
offset, from the axis, of the driving bunch trajectory.
For $s>L$ we can drop the scalar potential term
(it must be zero when there is no charge in the cavity), and
 the result can be written in the form\cite{bane}
\begin{equation}
{\bf W}(s)=\sum_\lm{c\over 2U_\lm\omega_\lm Lx_0}\Im{\rm m}\left[
V_\lm^*\nabla_\bot V_\lm \,e^{i\omega_\lm s/c}\right]
\quad\quad[s>L]\ ,
\label{eqcondc}
\end{equation}
with 
\begin{equation}
V_\lm=\int_0^L dz\,a_{\lm z}(z)e^{i\omega_\lm z/c}\quad.
\end{equation}
Note that the arbitrary constants associated
with the parameter ${\bf a}_\lm$ in the numerator and the denominator of 
Eq.~\ref{eqcondc} cancel.
Note also that---to the same arbitrary constant---$|V_\lm|^2$
 is the square of the voltage lost by
the driving particle to mode $\lm$ and 
$U_\lm$ is the energy stored in mode $\lm$.

Consider now the case of 
a cylindrically-symmetric, multi-cell accelerating cavity, and let us limit 
our concern to the effect of the dipole modes of such a structure.
We will allow the charges to move on an arbitrary, zig-zag path in the $x-z$ plane
that is close to the axis, and for which the slope is everywhere small
(so that $\nabla_\bot\sim \partial/\partial x$).
For dipole modes in a cylindrically-symmetric, multi-cell
accelerator structure, it can shown that
the synchronous component of $a_{\lm z}$ (the only component that, on
average, is important) can be written in the form $a_{\lm z}=xf_\lm(z)$
(see {\it e.g.} Ref.~\cite{Trans}).
Then Eq.~\ref{eqcondc} becomes
\begin{eqnarray}
W_x(s)&=&\sum_\lm{c\over 2U_\lm\omega_\lm Lx_0}\times\label{eqconde}\\
& &\hspace*{-11mm}\times\Im{\rm m}\left[e^{i\omega_\lm s/c}
\int_0^L dz^\prime\, x(z^\prime)f_\lm(z^\prime)e^{-i\omega_\lm z^\prime/c}
\int_0^L dz\, f_\lm(z)e^{i\omega_\lm z/c}
\right]\ [s>L]\ .\nonumber
\end{eqnarray}
Note that this equation can be written in the form:
\begin{equation}
W_x(s)= \sum_\lm 2k^\prime_{\lm}\sin\left({\omega_\lm s\over c}+\theta_\lm\right) 
\quad\quad[s>L]\ ,
\label{eqcondb}
\end{equation}
with $k^\prime_{\lm}$ a kind of kick factor and $\theta_\lm$ the phase
of excitation of mode $\lm$.
Note that in the special case
where the particles move parallel to the axis, at offset $a$,
$k^\prime_{\lm}=k_\lm=c|V_\lm|^2/(4U_\lm\omega_\lm a^2L)$,
 the normal kick factors for the structure,
and $\theta_\lm=0$.
For this case it can be shown that Eq.~\ref{eqcondb} is valid for
all $s>0$\cite{bane}.
Finally, note that, for the general case,
Eq.~\ref{eqcondb} can obviously not be  extrapolated down to $s=0$,
since it implies that $W_x(0)\not=0$, which is nonphysical,
since a point particle cannot kick itself transversely.
To obtain the proper equation valid down to $s=0$ we
would need to include the scalar potential term that was dropped in going
from Eq.~\ref{eqconda} to Eq.~\ref{eqcondc}.  

To estimate the wakefield associated with very small, random cell-to-cell
misalignments in accelerator structures we assume that 
we can use the mode eigenfrequencies and eigenvectors of the error-free
structure.
We obtain these from the circuit program.
Then to find the kick factors we replace
$x(z)$ in the first integral in Eq.~\ref{eqconde}
by the zig-zag path representing the negative of the cell misalignments, a path
we generate using a random number generator.
The normalization factor $x_0$ is set to the rms of the misalignments.

In Ref.~\cite{BL} this method is used to estimate the wake at
the bunch spacings in the S-band injector linacs of the JLC/NLC.
How can we justify this?
For example, for
the $3\pi/4$ S-band structure, one possible bunch spacing
is only 42~cm whereas the whole structure length
$L=4.46$~m. Therefore, in principle,
Eq.~\ref{eqconde} is not valid until the 11$^{\rm th}$ bunch spacing.
We believe, however, that the scalar potential fields will not
extend more than one or two cells 
behind the driving charge (the cell length is 4.375~cm), and
therefore this method will be a good approximation at all bunch
positions behind the driving charge.
This belief should be tested in the future by repeating
the calculation,
but now also including the contribution from scalar potential terms.

\begin{figure}[htb]
\centering
\epsfig{file=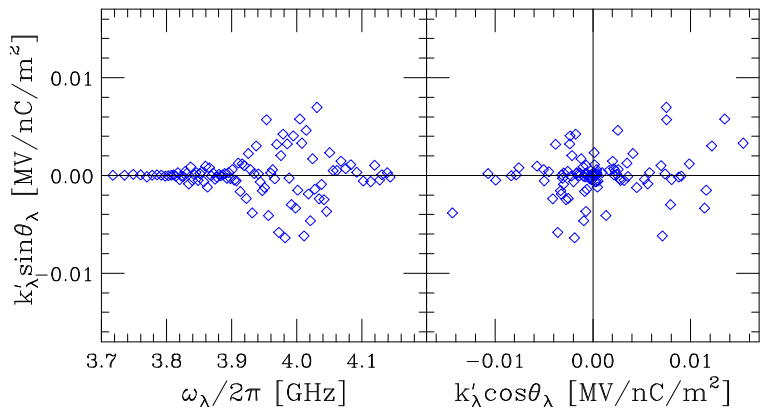, width=10cm}
\caption{
The kick factors and phases of the modes for a cell-to-cell
misalignment example. The structure is the optimized $3\pi/4$ S-band structure
for the injector linacs of the NLC (see Ref.~\cite{BL}).
}
\label{fiscatter}
\end{figure}

In Fig.~\ref{fiscatter} we give a numerical example. 
Shown, for the optimized $3\pi/4$ S-band structure 
for the injector linacs of the NLC(see Ref.~\cite{BL}),
are the kick factors and the phases of the modes as calculated
by the method described here. Note that $\theta_\lambda$ is
not necessarily small.

\section*{Acknowledgments}

The authors thanks V.~Dolgashev for
carefully reading this manuscript.

\end{document}